\documentstyle[twoside,fleqn,espcrc2]{article}

\title{ High Temperature Limit of the $ N= 2 $ IIA  Matrix Model}
\author{Subrata Bal and  B. Sathiapalan 
\address{The Institute of Mathematical Sciences, CIT Campus, Madras - 600 113,  INDIA}
\thanks{Talk presented by Subrata Bal at LATTICE 2000}
\thanks{  email- subrata,bala@imsc.ernet.in}
}
\begin{document}

\begin{abstract}

	The high temperature limit of a system of two D-0 branes is 
investigated. The partition function can be expressed as a power series 
in $\beta$ (inverse temperature). The leading term in the high temperature 
expression of the partition function and effective potential is 
calculated {\em exactly}. Physical quantities like the mean square 
separation can also be exactly determined in the high temperature limit.
We comment on $SU(3)$ IIB matrix model and the difficulties to study it.

\vspace*{-7mm}

\end{abstract}

\maketitle

\newcommand{\bel}{\begin{equation}\label}
\newcommand{\f}{\frac}
\newcommand{\bee}{\begin{equation}}
\newcommand{\ee}{\end{equation}}
\newcommand{\br}{\begin{eqnarray}}
\newcommand{\brr}{\begin{eqnarray*}}
\newcommand{\er}{\end{eqnarray}}
\newcommand{\err}{\end{eqnarray*}}
\newcommand{\pr}{\partial}
\newcommand{\non}{\nonumber \\}
\def\tr{\hbox{tr}} 
\def\th{\theta} 
\renewcommand{\thesection}{\arabic{section}}
\renewcommand{\thesubsection}{\thesection.\arabic{subsection}}
\renewcommand{\theequation}{\thesubsection.\arabic{equation}}

\section{Introduction}
\setcounter{equation}{0}

The study of string theory at finite temperature 
 has received renewed attention recently. In finite temperature 
 the number density of string states increases exponentially 
 with energy or temperature $[ \rho_b(m) = m^{-\f{25}{2}} \exp(4 \pi 
 \sqrt{\alpha'}m) ]$. So, partition function diverges for $T> T_H$, 
 where $T_H$ is the Hagedorn temperature. In perturbative string theory
 it is not clear wheather $T_H$ is limiting temperature or 
 phase transition temperature, so non-perturbative formulation
 like matrix model is needed. 
In a recent paper one of us \cite{BS1} attempted to elucidate
the nature of the Hagedorn transition \cite{H}  using the matrix model and found
 similarities with the deconfinement transition in gauge theories.
This was also investigated in a subsequent paper using the AdS/CFT
correspondence \cite{KRBS}. It is clear that recent developments in
 non-perturbative string theory or M-theory 
\cite{BFSS,IKKT,FKKT,AIKT} have some bearing
on our understanding of the high temperature behavior of strings.
For all these reasons the study of matrix models
 at high temperature is worthwhile.

A related model of D-instantons,
the IKKT matrix model \cite{IKKT}, which is 0+0 dimensional 
has been solved exactly for N=2 \cite{TSU}. The D-0-brane action that we are
interested in, is a quantum
mechanical one (i.e. 0+1 dimensional). However after compactifying
the Euclideanised time, if one takes the high temperature limit, it
reduces to a 0+0 dimensional model. There is thus a hope of solving this model
 order by order in $\beta$ but {\em to all orders in $g$} using the
 same techniques as \cite{TSU}. One can
then calculate physical quantities such as the mean square
separation of the D-0-branes - a measure of the size of the bound
states.  This is what is attempted in this paper.
We obtain the leading behavior in $\beta$.
We can also estimate, 
the corrections to the leading result. The noteworthy feature being
that
each term is 
{\em exact in its dependence on the string coupling constant }. 
For details and references see \cite{BALS1}.

\section{High Temperature $SU(2)$ matrix Model}\label{himm} 

\subsection{The action}\label{ac} 
\setcounter{equation}{0}

\noindent
The BFSS Lagrangian is given by 10  dimensional SYM lagrangian reduced
in $(0+1)$ dimension 
\br
L={1\over 2g l_s} \tr \left[   \dot{ X^i}\dot{ X^i} + 2 i \th
 \dot{\th} - {1\over 2 l_s^4}[X^\mu,X^\nu]^2 
\right. \non \left.
-
\f{2}{l_s^2} \bar{\th} \gamma_\mu [\th ,X^\mu ] 
- \f{i}{l_s^2} [ X^0, X^i] \dot{X}^i \right]
\label{lag1}
\er
where $i = 1,...,9; \mu = 0,..., 9$
; $X^\mu$ and $\th$ are $N $ x  $ N$ hermitian matrices.
$X^\mu$ is 10  dimensional vector and $\th$ is 16 component Majorana-Wyel 
spinor in 10 dimension.
We write $X^\mu, \th$ in terms of the group generators $T^a$
\bee
X^\mu = \sum_{a =1}^{N^2-1} T^a X^\mu_a,
~~~~~~~~
\th =\sum_{a=1}^{N^2-1} T^a \th_a 
\ee

$X^\mu_a, \th_a $ are real fields.
For $N = 2$, $T^a = {1 \over 2} \sigma^a $,
$\sigma^a$ are the hermitian Pauli matrices.

\noindent
If we Euclideanize ($t \rightarrow it, X^0 \rightarrow  i X^0$)
 and compactify time on a circle of circumference  $\beta$,
the action becomes 

\bee  
S = i \int_0^\beta L dt 
\ee  

\noindent
Considering the field boundary conditions

\[ X^\mu(0) =  X^\mu(\beta) , ~~~~~~
 \th(0) = - \th(\beta) \]

\noindent
we can expand the fields $X^\mu, \th $
in modes as

\[
 X^\mu_a(t) = \sum_{n = - \infty}^{\infty} 
 X^\mu_{a,n} e^{\f{2 \pi i n}{\beta} t},
~
 \th_a(t)  = \sum_{r = - \infty}^{\infty} 
 \th_{a,r} e^{\f{2 \pi i r }{\beta} t}
\]
$n,(m,l,p)$ are integers and $r,(s)$ are  half-integers .

\noindent
So, the action reduces to

\bel{acn6}
  S =  S_{b,free} +  S_b +  S_f
\ee
where,  $S_{b,free}$, $S_b$ and  $S_f$ are the  
free  bosonic, bosonic and the fermionic terms

\bel{acnb0}
S_{b,free} = \f{i}{4 g} \left\{ - \sum_{n= - \infty}^{\infty}
\f{4 \pi^2  n^2}{\beta^2}
X^i_{a,n} X^i_{a,-n}
\right\}
\ee
\br
S_{b} = \f{i}{8 g} 
\left\{
\beta
 \sum_{\stackrel{n,m,l,p = - \infty,}{ n+m+l+p=0}}^{\infty}
 X^\mu_{a,n} X^\nu_{b,m} X^\mu_{a,l}
 X^\nu_{b,p} 
\right. \non
\left.
-
\beta
 \sum_{ { \stackrel{n,m,l,p = - \infty,}{n+m+l+p=0} 
 }}^{\infty}
 X^\mu_{a,n} X^\nu_{b,m} X^\mu_{b,l}
 X^\nu_{a,p}
\right. \non
\left.
+ n \pi \epsilon^{abc}
 \sum_{ n + l + p = 0} 
X^0_{a,l} X^i_{b,m} X^i_{c,n}
\right\}  \label{acnb02}
\er
\br
S_{f} = \f{i}{4 g} 
\left\{
-2 i \beta 
 \sum_{\stackrel{r,s,l = - \infty,}{\footnotesize {\em  s+r+l = 0}}}^{\infty}
 \th_{a,r} \gamma_0 \gamma_\mu \th_{b,s}
 X^\mu_{c,l} 
\right. \non
\left.
 +
\sum_r
 4 \pi r
 \th_{a,r}  \th_{a,-r}  
\epsilon^{abc}
\right\} 
\label{acnf1}\er  

$\left(X^\mu_{a,n}\right)_{n \neq 0}$ and $X^\mu_{a,0}$ are of the order 
$\left( \f{\sqrt{\beta}}{n} \right)$ and (1). In $\beta \rightarrow 0$
limit, the zero-modes of $X^\mu$ will contribute in the leading order 
to $Z$. So, we can re-write

\br
S_{b} = \f{i \beta}{8 g} 
\left\{
 X^2_{a,0} X^2_{b,0} - \left( X_{a,0}.X_{b,0}\right)^2 
\right\}  \label{acnb01}
\er
\br
S_{f} = \f{i}{2 g} 
\left\{
- i \beta 
 \sum_r
 \th_{a,r} \gamma_0 \gamma_\mu \th_{b,-r}
 X^\mu_{c,0} 
\right. \non
\left.
 +
\sum_r
 2 \pi r
 \th_{a,r}  \th_{a,-r}  
\epsilon^{abc}
\right\} 
\label{acnf}\er  

\noindent
The partition function is $Z = \int e^{-iS} $

\subsection{Pfaffian}
\setcounter{equation}{0}

\noindent
Following \cite{TSU}, we rotate $X^\mu_{c,0}$ by a Lorentz transformation so that only $X_{c,0}^0$,
$X_{c,0}^1$ and $X_{c,0}^2$ are nonzero. We take the representation of the 
Gamma matrices, in which

\bel{gammac}
\gamma_0 = i \sigma_2 \otimes 1_8, ~~~ 
\gamma_1 =  \sigma_3 \otimes 1_8 , ~~~
\gamma_2 = - \sigma_1 \otimes 1_8 
\ee

\noindent
Integrating the fermionic fields we get the pfaffian,

\[
Z_f = g^{-24} \left( O(1) + O(\beta^\f{3}{2} g^\f{1}{2})
 + O( \beta^\f{9}{4} g^\f{1}{4} ) + .... \right)
\]

\noindent
the above expression has  $SO(3)$ symmetry 
in spinor indices, and $SO(2,1)$ symmetry in the vector indices.
The $O(1)$ term gives the free fermionic contribution. Note that it is 
temperature independent as the Hamiltonian is identically 
zero for free fermions in $0+1$ dimensions.

\subsection{Free Bosonic sector}
\setcounter{equation}{0}

\noindent
As the higher modes interacting terms do not contribute to the leading 
order, we can decouple the free bosonice and interacting terms. 
Integrating the bosonic sector and  regularizing 
, we get 

\bee
Z_{free} = \left( \beta g \right)^{-27}
\ee

\label{lead}\subsection{ Leading and Non-leading Interaction Terms.}
\setcounter{equation}{0}

\noindent
The terms $X^0_{a,l} X^i_{b,m}X^i_{c,n}$
and $X^i_{a,n} X^i_{a,-n}$ in the original action are not Lorentz invariant. 
However, these terms do not contribute to the partition function in leading 
order . The action with only the zero modes
has Lorentz invariance. Hence, as long as we are interested in leading 
order contribution only, we can work with and assume Lorentz
invariance and consider the parametrisation 

\bel{param}
X_{1,0} = ( x_1, \vec{r}_1 ), ~~
X_{2,0} = ( x_2, \vec{r}_2 ),  ~~
X_{3,0} = ( l, 0 ) 
\ee

\noindent
At this stage, we can find the temperature dependence of the partition function 
and the mean square separation of two D-0 branes from a simple scaling argument. 
This scaling argument in fact applies to $SU(N)$ also. To see this we
note that the leading order
$i.e.$ the zero mode bosonic contribution of the partition function comes 
from the $[X^\mu,X^\nu]^2 $ term in the Lagrangian 
(\ref{lag1}). And this term is  Lorentz invariant and hence just as
in $SU(2)$ case can use Lorentz invariance while  calculating the
leading order contribution to the partition function from this term. 
Therefore in  $SU(N)$ case also we can use a parametrisation  similar 
to \ref{param}

\[
X_{i,0} = ( x_i, \vec{r}_i ), ~~1\leq i \leq N^2 -2 ;~~
X_{N^2-1,0} = ( l, 0 )
\]

Under the above 
parametrisation
${1\over 2 }[X^\mu,X^\nu]^2 $
will be  homogeneous in
$l, r_i, x_i$,  $0 < i \leq N^2 -2$  and of order 4. So, in general 
for any $N$ , 
we need to scale these variables by  $\beta^{-\f{1}{4}} g^{\f{1}{4}} $
to scale out the $\beta$ from the exponent. 
And the temperature dependence of $\langle l^2 \rangle$
will be $\beta^{-\f{1}{2}} g^{\f{1}{2}}$ in the leading order.
Under the above scaling the measure in $Z_0$  will pick up
a $\beta^{-\f{15}{2}} g^{\f{15}{2}}$ factor for $SU(2)$
, which comes from ($3 \times 10$) 
$X^\mu_{a.n}$.  In general for $SU(N)$ in $D$ dimension there will be 
$ D (N^2 - 1)$ $X^\mu_{a,n}$ in the measure. So, the partition function $Z_0$ 
has temperature dependence  $\beta^{-\f{D (N^2 - 1)}{4}} g^{\f{D (N^2 - 1)}{4}}$.
And $Z_{free}$ will be proportional to $ (\beta g)^{(D-1)(N^2 -1)}$.

\noindent
Now we evaluate the partition function for this action.
 
\bee
Z_0 =
\f{2^{34} 15 ~g^{\f{15}{2}} \pi^5 }{\beta^{\f{15}{2}}}
\int_{-\infty}^\infty dl  I_0(l)
\ee

\noindent
where 
\br
 I_0(l)=
24576 l^{-7} - 2752512 l^{-11} + ....
\label{largelz}
\er
In large $l$ limit and
\br
I_0(l) =
 \sqrt{8 \pi } l^{7}
- \f{256}{5} l^9
+ \left( \sqrt{\f{\pi}{8} } - \f{256}{3} \right) l^{11}
+.... 
\label{smalllz}
\er
In small $l$ limit.

\noindent
Hence, $Z$ converges for both large $l$ and small $l$ and 
$I_0(l)$ is non-singular and independent of $\beta$. 
\noindent
So, $Z_0 $ has a temperature dependence of $T^{\f{15}{2}}$.

Combining the free bosonic, fermionic and the bosonic parts, we can write 
the $\beta$ dependence as

\bee
Z= 
\beta^\f{39}{2}
\left( O(1) + O(\beta^\f{3}{4}) 
+ O(\beta^{\f{9}{8}})
+ ...\right)
\ee

\subsection{Effective Potential \& Mean-square Separation of the D-0 branes.}
\setcounter{equation}{0}

For high temperature we have evaluated the partition function both for large
and small $l$.
Up to leading order the effective
potential between two D-0 branes is proportional
to $- \log l$ and $ \log l$ for small and large $l$
(\ref{largelz},\ref{smalllz}).  We can see that
the potential
increases at both $l$ ends,  though we can not clearly see the nature of the
potential in the intermediate region but we can
conclude that the potential is a confining potential and binds the
 D-0 branes.

We can 
identify
$l$ as one of the spatial components and hence as the separation between 
two D-0 branes.
We calculate the mean square separation of two D-0 branes.

\bel{msq}
 \left\langle l^2 \right\rangle = 6.385 \left(\f{\beta}{ g }\right)^{-\f{1}{2}}  \ee

\noindent
If we assume high temperature expression has a finite radius of convergence,
we can conclude that 
the mean square separation is finite for finite temperature. This implies that there is a confining
 potential that binds the D-0 branes. As argued earlier the scaling argument that gives the $\beta$ and $g$ dependence in \ref{msq} is valid for {\em all N }. So we can 
conclude that $ \langle l^2 \rangle \approx \sqrt{\f{g}{\beta}} $ for all $N$.

\section{Attempt to study $SU(3)$ IIB matrix model}

As a step towards understanding the large $N$ matrix model,
after studying $SU(2)$ matix model, we try to study the $SU(3)$
matrix model. But, $SU(3)$ has 8 generetors, so the $SO(10)$ 
rotation will leave atleast 8 of the $X^\mu$ non-vanishing resulting 
non-zero coefficient for 8 $\gamma$ matrices in the expression
$\gamma_\mu X^\mu$, unlike the $SU(2)$ case, where we had 3 generators
and 3 non-zero coefficients. In $SU(3)$ case, we have to work with
$16 \otimes 16 ~ \gamma$ matrices, unlike $SU(2)$, where we choose 
the suitable form as in \ref{gammac}. Pfaffian for $SU(3)$ is determinant of
a $128 \otimes 128$ matrix. Moreover, we have at least 24 parameters
to integrate to calculate $Z$. All these makes $SU(3)$ difficult to study.
But we can $X^\mu_a T^a= X^\mu_i T^i + X^\mu_8 T^8 + X^\mu_l T^l$ and
$\th_a T^a= \th_i T^i + \th_8 T^8 + \th_l T^l$ and look the parturbative 
solution in the limit $X^\mu_8, X^\mu_l \rightarrow 0$, which will give 
the $SU(3)$ partition function as $SU(2)$ partition function plus 
modifications. We are working on that presently and will report soon.

\section{Conclusion}
\setcounter{equation}{0}

In this paper we have attempted to study a system af two d0 branes at 
distance $l$ kept in  high temperature.
 The leading nontrivial term of the partition function 
has been calculated exactly. 
 The non-leading terms can
also be 
systematically calculated 
although we haven't attempted to work them out in this paper. From a scaling argument 
we have also determined the $\beta$ and $g$ dependence of the leading term for any $N$. This 
complements the work 
of \cite{AMB}, where the one loop partition function was calculated with the 
entire $\beta$ dependence.
We find that $ \langle l^2 \rangle \propto
\sqrt{\f{g}{\beta}} $ (eqn. \ref{msq}) (true for any $N$), 
the finiteness
of which shows that there must be a potential between D-0 branes that binds
them. In \cite{BS1,AMB}  also a logarithmic and attractive potential
were found. The present calculation being exact in $g$ is valid at
all distances.  Thus unlike in \cite{BS1,AMB},
the (finite temperature) logarithmic potential found here is 
attractive at long distances
and repulsive at short distances so it has a minimum at non-zero
separation.
In \cite{BS1} it
was found that at high temperatures, 
the configuration with all the D-0-branes clustered
at the origin $i.e.$ with the zero separation,
had lower free energy than the one where they were
spread out. However, that was a large $N$ calculation and also restricted
to one loop. 
It is therefore possible that more  exact calculation will resolve this 
issue.

\end{document}